\documentclass[aps,prb,twocolumn,amsmath,amssymb,gallery,footinbib,superscriptaddress,floatfix]{revtex4-2}
\usepackage{graphicx}
\usepackage{dcolumn}
\usepackage{bm}
\usepackage{hyperref}
\usepackage{stackrel,amssymb}
\hypersetup{colorlinks=true, linkcolor=blue, citecolor=blue, urlcolor=magenta, pdftitle={On the search of displacive chiral phase transitions}, pdfauthor={F. Gomez-Ortiz and E. Bousquet}}
\usepackage{fixltx2e}
\usepackage{bbding}
\usepackage{xcolor} 
\usepackage[normalem]{ulem}

\begin{document}
\preprint{APS/123-QED}
\title{Pathways to crystal chirality An algorithm to identify new displacive chiral phase transitions}
\author{Fernando G\'omez-Ortiz}
\email[Corresponding author: ]{fgomez@uliege.be}
\affiliation{Physique Théorique des Matériaux, QMAT, CESAM, Université de Liège, B-4000 Sart-Tilman, Belgium} 
\author{Aldo H. Romero}
\affiliation{Department of Physics and Astronomy, West Virginia University, Morgantown, WV 26505-6315, USA}
\author{Eric Bousquet}
\email[Corresponding author: ]{eric.bousquet@uliege.be}
\affiliation{Physique Théorique des Matériaux, QMAT, CESAM, Université de Liège, B-4000 Sart-Tilman, Belgium} 
\date{\today}
\begin{abstract}
We present an algorithm that integrates pseudosymmetry search with first-principles calculations to systematically identify achiral parent structures and establish potential chiral displacive transitions linking them to their corresponding chiral phases within the 22 enantiomorphic space groups. This approach enables a robust exploration of structural relationships, offering new insights into symmetry-driven properties.
Our workflow streamlines the discovery of displacive chiral phase transitions driven by soft phonon modes, providing insights into the mechanisms of structural chirality in inorganic materials.
We apply this methodology on the chiral phases of TeO$_2$, Na$_2$SeO$_9$, Sr$_2$As$_2$O$_7$, As$_2$O$_5$, Rb$_2$Be$_2$O$_3$, and CaTe$_2$O$_3$. 
Demonstrating that some do not have a minimal supergroup that allows for an achiral phase; some can have a minimal supergroup, still, no unstable phonon mode exists in the achiral phase; and somewhere the minimal supergroup exists with a soft phonon mode connecting the identified achiral phase and the chiral phase through small continuous displacements.
\end{abstract}
\maketitle
\section{Introduction}
\label{sec:introduction}
Chirality in periodic solids is characterized by structures belonging to space groups that include only operations of the first kind~\cite{Nespolo2018,souvignier2003enantiomorphism, bousquet2025,fecher2022,avnir2024}, i.e., those which map a right (left)-handed coordinate system to another right (left)-handed system. 
This fundamental property distinguishes chiral structures from their achiral counterparts~\cite{sohncke1879entwickelung,le1998concept}. 
The concept of chirality in crystallography extends beyond mere point group symmetry and encompasses the complete three-dimensional periodicity of the crystal lattice~\cite{flack1999absolute}. 
Chiral space groups, also known as Sohncke groups, comprise 65 of the 230 crystallographic space groups and are characterized by the absence of improper symmetry operations such as inversion centers, mirror planes, and rotoinversion axes~\cite{sohncke1879entwickelung,flack2003chiral,Nespolo2018}. 
This inherent asymmetry gives rise to unique physical properties in chiral materials, including optical activity~\cite{Arago1811,pasteur1848relations} or nonlinear optical effects~\cite{verbiest1998strong}, which have significant implications in various fields ranging from materials science to pharmaceutical research~\cite{barron2004raman}.
Recently, the study of chiral effects in crystals has gained considerable attention, particularly concerning phenomena such as chiral phonons~\cite{Zhang-15,zhu-18,romao2024,Wang2024}, which are phonons that exhibit rotational motion (clockwise or counterclockwise) that are locked with the particle’s momentum. 
Another area of interest is spin selectivity~\cite{Dalum-19}, where one spin state is favored over another when electrons pass through chiral magnetic structures. 
Additionally, a recent research focus is polar and magnetic skyrmions~\cite{Junquera-23}, topologically protected textures of spin or electric dipoles.
There is a growing interest in exploring topological states, such as those associated with Kramers-Weyl fermions~\cite{Chang-18}, alongside the emerging field of altermagnetism~\cite{Vsmejkal-22, Mazin-22}. Chiral crystal structures, characterized by the absence of inversion symmetry and mirror planes, are particularly promising in this context. These unique structures can give rise to distinctive magnetic properties, positioning them as natural candidates for altermagnetic phenomena~\cite{Vsmejkal-22} while also profoundly influencing the electronic structure and band dispersion~\cite{sanchez2019topological}.

Theoretically, irreducible representations (irreps) that satisfy the group theoretical conditions for chiral objects formally exist~\cite{Hlinka-14}, as these irreps inherently possess characteristics that align with the symmetry requirements for chirality. This implies that symmetry-based frameworks can differentiate between chiral and achiral states.
Consequently, phase transitions involving spontaneous symmetry breaking, characterized by an order parameter that distinguishes between chiral and achiral phases, also exist and should be observed in nature, as the order parameter is a physical quantity that changes during the transition, reflecting the emergence of chirality~\cite{bousquet2025}.
The crystallization process within a chiral symmetry group plays a decisive role in determining the material's chirality and its associated properties, making it a crucial area of investigation~\cite{uwaha2016introduction,katsuno2016mechanism, viedma2005chiral}. 
Therefore, understanding such structural phase transitions from an achiral to a chiral phase is essential for identifying the material's final symmetry and its electronic, magnetic, and optical properties~\cite{wagniere2011chirality}.

A fascinating case occurs when such transition is driven by an unstable phonon mode that transitions from an achiral to a chiral phase, as recently proposed for K$_3$NiO$_2$~\cite{fava2025-KNO}. 
In this case, the phase transition is displacive or order-disorder (i.e., not reconstructive), and a group-supergroup relation between the two phases exists, which also means it preserves the internal connectivity and does not break any bonds~\cite{Dove-97}.
However, despite significant theoretical and experimental efforts in this field, only a very few number of candidate materials exhibiting transitions between chiral and achiral phases have been identified to date~\cite{bousquet2025}:  MgTi$_2$O$_4$~\cite{Isobe2002,Schmidt2004}, K$_3$NiO$_2$~\cite{fava2025-KNO}, CsCuCl$_2$~\cite{Schlueter-66}, Na$_2$Ca$_2$(SiO$_3$)$_3$~\cite{maki-68,fischer-87, ohsato-90} or Ag$_4$P$_2$O$_7$~\cite{yamada-83}.  Although Pb$_5$Ge$3$O${11}$ has also been reported to exhibit an achiral-to-chiral phase transition through a polar soft mode~\cite{Fava-23}, it is excluded from the present discussion as we focus specifically on the 22 enantiomorphic crystal space groups.

Therefore, several fundamental questions arise: Are there intrinsic physical or chemical barriers inhibiting the existence of such phase transitions naturally occurring in nature, or have they been overlooked due to limitations in our search methodologies~\cite{wachter2016exceptional, brock2016high}? 
Could a more meticulous and systematic exploration of diverse crystal structures leverage advanced computational screening techniques~\cite{jain2016computational,butler2016computational} uncover additional candidates for chiral phase transitions? 
Do we possess the appropriate theoretical frameworks and experimental techniques to effectively investigate and characterize these potential transitions? The answers to these questions could significantly impact our understanding of chirality in condensed matter systems and potentially lead to the discovery of novel materials with unique properties~\cite{wagniere2011chirality}.

In this study, we propose a systematic methodology for identifying achiral parent structures from known chiral crystalline phases belonging to the 22 enantiomorphic space groups and capable of exhibiting displacive phase transitions. We employ both rigorous symmetry analysis and first-principles calculations.
For simplicity and clarity, we restrict our focus to non-magnetic crystals displaying chiral enantiomorphic space groups as magnetic chirality can be at play as well~\cite{bousquet2025}. 
We use group-theoretical approaches to derive the parent structure from the atomic positions of the low-symmetry chiral phase. 
While illustrated here on a proof-of-concept set of six compounds, the methodology is well suited for future high-throughput implementation, offering the potential to uncover novel candidates for chiral transitions.
Once the parent achiral structure has been identified relying on symmetry arguments, subsequent density functional theory (DFT) calculations are conducted to compute phonon instabilities and check the existence of soft modes connecting the achiral reference and the chiral structures.
By presenting this framework, we aim to stimulate the scientific community to identify and investigate novel materials exhibiting chiral transitions. This could potentially lead to discoveries with implications for applications such as optoelectronics, spintronics, and nonlinear optics.


The article is organized as follows: First, we will present the systematic procedure to obtain high-symmetry parent structures belonging to achiral space groups for a given low-symmetry chiral phase. 
In the following sections, we will apply this technique to some candidates within the 22 enantiomorphic space groups.
Our methodology helps identify crystals that can exhibit displacive chiral phase transitions from the combination of symmetry and first-principles analysis.

\section{Results}
\subsection{Finding an achiral reference for a given chiral crystal}

In this section, we will outline the necessary steps to obtain an achiral reference structure starting from the atomic positions of a low-symmetry chiral crystal.
We will restrict ourselves to crystals from the 22 enantiomorphic space groups, but the methodology could be applied to all Sohncke groups.
To this end, we will conduct a symmetry analysis following the successful methodology already employed in the determination of new displacive ferroelectric materials~\cite{Kroumova-00,Kroumova-02,Capillas-04} or high-temperature structural transitions of $Pnma$ or $P2_12_12_1$ symmetries~\cite{Igartua-99,Igartua-96}. This method acts as a screening test; if no candidate structures are found, we can confirm that the crystal in question cannot be connected to an achiral reference through phonon distortions.
However, if candidate structures are identified, further first-principles stability calculations and phonon analysis must be performed to confirm the presence of an unstable chiral phonon that can lower the energy of the achiral phase to the chiral one.

Suppose we have a chiral crystal belonging to a particular group $\mathcal{G}$ occupying the atomic positions $\mathbf{r}_i$. 
If we require this material to show a displacive phase transition from a higher symmetry achiral reference, then the atomic positions need to comply
\begin{equation}
    \mathbf{r}_i=\mathbf{r}^0_i+\mathbf{u}_i,
    \label{eq:pseudosymmetry}
\end{equation}
where $\mathbf{r}^0_i$ are the positions of the atoms in the parent structure belonging to a higher symmetry achiral space group $\mathcal{H}$, and $\mathbf{u}_i$ are the set of individual atomic displacements that drive the displacive phase transition. Moreover, the enantiomorphic spacegroup $\widetilde{\mathcal{G}}$ would present a similar relation $\widetilde{\mathbf{r}}_i=\mathbf{r}^0_i+\widetilde{\mathbf{u}}_i$ where $\mathbf{u}_i$ and $\widetilde{\mathbf{u}}_i$ are related by an improper rotation.

In this context, the original structure exhibits pseudosymmetry, denoted as $\mathcal{H}$~\cite{Kroumova-01}. The atomic positions, $\mathbf{r}^0_i$, may correspond to the high-symmetry, high-temperature phase within the phase diagram, which, in our case, serves as the achiral reference structure.
Therefore, finding crystalline structures belonging to supergroups $\mathcal{H}$ complying with the restrictions imposed by Eq.~\ref{eq:pseudosymmetry} can give candidates for displacive phase transitions towards the low-symmetry phase under consideration.
Indeed, such methodology has proven effective in predicting displacive ferroelectric phase transitions~\cite{Kroumova-00,Kroumova-02,Capillas-04} or high-temperature structural transitions~\cite{Igartua-99,Igartua-96}.

However, searching among all the possible supergroups $\mathcal{H}$ of a given spacegroup $\mathcal{G}$ is unnecessary. 
A better strategy is to start screening among the minimal supergroups of $\mathcal{G}$. Indeed any group-supergroup relation ($\mathcal{G}<\mathcal{H}$) can be decomposed by a chain of minimal supergroups $\mathcal{G}<\mathcal{K}_1<\mathcal{K}_2...<\mathcal{H}$~\cite{Dummit-03}. Therefore, finding no candidates among the minimal supergroups of $\mathcal{G}$ guarantees the absence of displacive phase transitions for the given crystalline structure. Conversely, if candidates are identified using the minimal supergroup approach, further searches can be conducted following the same strategy.

The search of minimal supergroups of $\mathcal{G}$ conveying with Eq.~\ref{eq:pseudosymmetry} can be done with the tool {\textit{Pseudosymetry search}}~\cite{Capillas-11} of the Bilbao Crystallographic server~\cite{Aroyo-06,Aroyo-06.2} which not only accounts for the symmetry of the initial subgroup but also takes into account the information of the Wyckoff positions occupied by the crystal to provide a list of consistent candidates. 
First, the determination of all minimal supergroups of a given space group can be performed by inverting the data on the maximal subgroups of space groups available in volume A1 (Symmetry Relations Between Space Groups) of the International Tables for Crystallography~\cite{Wondratschek2-10}.
Then, splitting the Wyckoff positions is considered an increase of site-symmetry may cause atoms belonging to distinct special positions in the subgroup to become equivalent in the higher-symmetry supergroup~\cite{Wondratschek-93}.
Hence, we can apply this method to the chiral crystals reported in the International Crystal Structure Database (ICSD)~\cite{Zagorac-19} to find novel structures displaying displacive chiral phase transitions.

Once the parent achiral reference has been determined, subsequent DFT calculations are required to complement the symmetry analysis. First, a relaxation of the atomic structure $\mathbf{r}^0_i$ is performed, followed by a phonon calculation to identify a potentially unstable phonon. 
These phonon modes are then analyzed to determine whether any of the instabilities could drive a transition to a chiral phase. Finally, the eigendisplacements associated with the unstable modes are condensed in the direction of the chiral phase to confirm the connection with the structure reported in the databases.

\begin{center}
  \begin{figure}[h]
     \centering
      \includegraphics[width=\columnwidth]{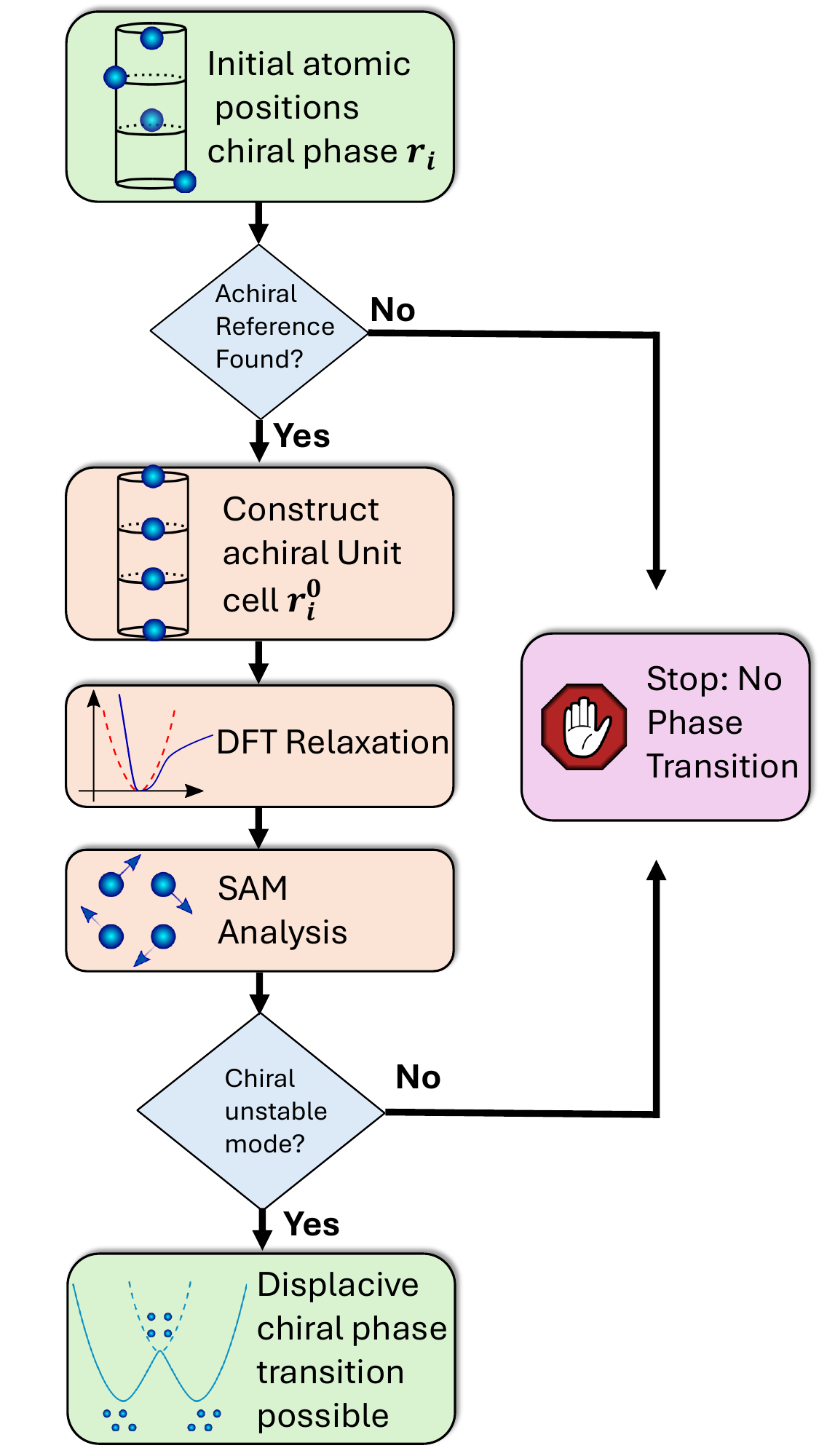}
      \caption{Algorithmic view of the methodology described in the article to search for new displacive chiral phase transitions.} 
      \label{fig:algorithm} 
  \end{figure}
\end{center}

We summarize the methodology workflow algorithm in Fig.~\ref{fig:algorithm}. As previously mentioned, we start with the atomic positions of the chiral phase and perform a pseudosymmetry search to identify potential achiral reference structures. The algorithm terminates if no compatible achiral reference structure is found, as a phase transition would be impossible. 
However, we proceed with DFT relaxations if an achiral reference is identified. 
Symmetry-adapted mode (SAM) analysis follows this step to identify at which special $k$-points an irrep can drive the high-symmetry structure to the low-symmetry chiral structure. Hence, only phonon calculations at those special $k$-points should be done, which helps to save computational time by avoiding doing the full phonon dispersion curve.
A phonon calculation is then conducted at those $k$-points where we know that an irrep can give the achiral to chiral phase transition if phonon mode instability with this irrep is present.
If no instabilities are detected, the algorithm stops; if instabilities are present, the algorithm condenses these modes to determine whether any lead to the chiral phase. If successful, the material is added to the list of candidates for displacive chiral transitions, completing the process.

\subsection{The case of T\MakeLowercase{e}O\textsubscript{2}}

In this section, we apply our methodology to the interesting case of TeO$_2$ crystal. 
Among the three commonly reported polymorphs at ambient conditions~\cite{LECIEJEWICZ-61,Thomas-88},  this material has an enantiomorphic low-symmetry phase of either $P4_12_12$ or $P4_32_12$ space groups. 
To our knowledge, no parent structure has yet been reported for this compound at the bulk level. Still, a rutile $P4_2/mnm$ high symmetry phase has been reported recently when grown epitaxially on FeTe substrate~\cite{peng2022}.
\begin{figure*}[btp]
     \centering
      \includegraphics[width=14cm]{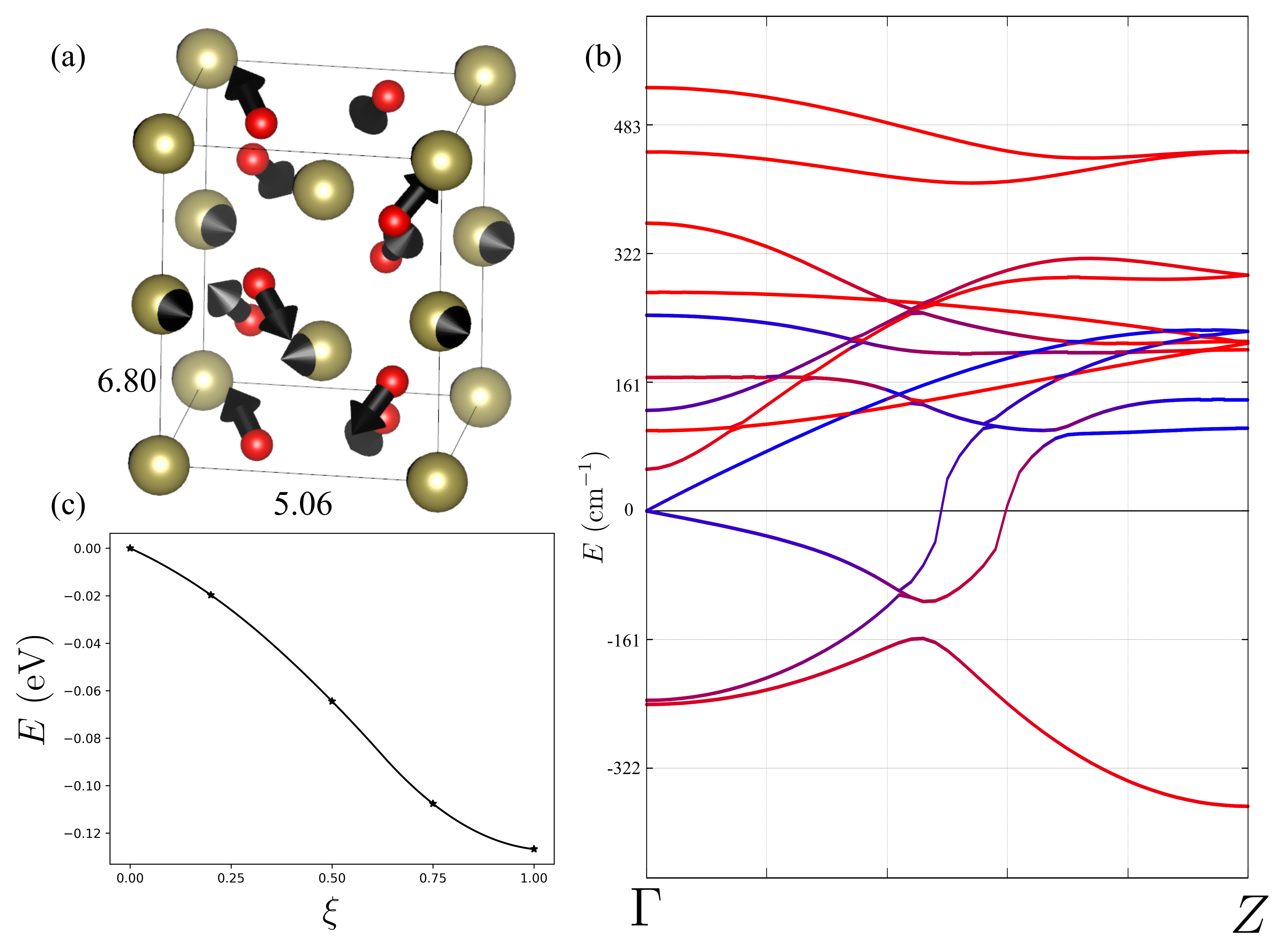}
      \caption{(a) Schematic representation of the TeO$_2$ crystal structure where atoms occupy the $P4_2/mnm$ high symmetry positions. 
      Arrows indicate the direction of the atom displacements that bring the system to the $P4_12_12$ chiral phase. 
      Gold and red balls represent Te and O atoms, respectively. (b) Calculated phonon band structure along the $\Gamma-Z$ path of the high symmetry phase of TeO$_2$. The color of the different branches are weighted according to the contribution of each chemical species to the dynamical matrix eigenvector (red for the O atom, and blue for the Te atom). (c) Energy surface when we condense the $Z_3$ unstable mode along the $(a,a)$ directions in normalized units.}
      \label{fig:structureTeO2} 
\end{figure*}
Our pseudosymmetry search predicts that the high symmetry $P4_2/mnm$ achiral structure can be linked to the chiral low-symmetry $P4_12_12$ (space group number 92) and $P4_32_12$ (space group number 96) phases through the $Z3$ irrep from the zone boundary $Z$ point, similarly to the case of K$_3$NiO$_2$\cite{fava2025-KNO}. The calculated structural data of the converged $P4_2/mnm$  structure computed with a $k$-mesh of 10$\times$10$\times$10 can be found in the appendix.

We can therefore study from DFT calculations whether a soft phonon mode is present at the $Z$ point of the $P4_2/mnm$ achiral phase of TeO$_2$. As shown in Fig.~\ref{fig:structureTeO2}(b), a double degenerate unstable phonon mode is encountered at $Z$, showing a frequency of $370i$ cm$^{-1}$.
The symmetry of this unstable mode corresponds to the irrep $Z_3$, which is the right one to link the high-symmetry rutile phase and the low-symmetry chiral phases.
Due to the double degeneracy of the $Z_3$ mode, different linear combinations $(a,b)$ can be constructed. 
Diagonal distortions $(a,a)/(-a,-a)$ and $(a,-a)/(-a,a)$ domains are associated to the $P4_12_12$ and $P4_32_12$ space groups respectively. 
Similarly to the case of K$_3$NiO$_2$, non-diagonal distortions $(a,b)$ drives the system to a $C222_1$ phase whenever $a\neq 0$ and $b\neq 0$ whereas $Cmcm$ phases are encountered when $a=0$ or $b=0$.
The condensation of the $Z_3$ unstable mode eigenvector give an energy gain of $\Delta E=-0.127$ eV per atom for the $(a,a)$ phase as shown in Fig.~\ref{fig:structureTeO2}(c). 
We also obtained that the $(a,b)$ case has the same energy gain as the $(a,a)$ case, so we can conclude that the case $a\ne b$ does not give a local minimum.
The $(a,0)$ case shows a higher energy ($\Delta E=0.06$ eV) than the $(a,a)$ case such that we can conclude that, through the $Z_3$ irrep, the chiral $(a,a)$ domains are those giving the lowest energy.

A SAM analysis of our relaxed chiral phase reveals that the distortion for the $P4_2/mnm$ phase decomposes into $Z_3$ and $\Gamma_1$ irreps where the amplitude of the $\Gamma_1$ mode (0.004~\AA per ion) is two orders of magnitude smaller than that of the $Z_3$ mode (0.093~\AA per ion).
The $\Gamma_1$ mode is a secondary SAM that appears through invariant coupling with the $Z_3$ unstable mode as this mode is not unstable at the zone center.

The $Z_3$ mode induces a slight in-plane displacements of the Te atoms, as well as both out-of-plane and in-plane displacements of the O atoms. However, the in-plane component of the O displacement is smaller than the out-of-plane one. As shown in the weighted phonon band structure of Fig.~\ref{fig:structureTeO2}(b) and the magnitude of the arrows in Fig.~\ref{fig:structureTeO2}(a) the distortion is mostly dominated by the O movements. 
Similarly to K$_3$NiO$_2$, the associated distortions produce a helical structure as shown in Fig.~\ref{fig:structureTeO2}(a).
In Fig.~\ref{fig:structureTeO2}(a), we show a schematic representation of the TeO$_2$ crystal and its chiral $Z_3$ $(a,a)$ distortion. 
Atomic positions correspond to the high-symmetry $P4_2/mnm$ phase, and the arrows indicate how the atoms move when transitioning from the achiral rutile phase to the chiral $P4_12_12$ phase. 
The handedness of the chiral phase has been computed following the helicity method described in Ref.~\cite{gomez2024pros,github}. This helicity is well-defined within the context of displacive phase transitions, allowing for a reference-independent comparison between different compounds. By normalizing its value to the number of atoms in the unit cell, the reference becomes intrinsically determined by the physical nature of the transition.
For TeO$_2$ a value of $7.38\cdot10^{-3}$ is found, which is similar to the one found in K$_3$NiO$_2$ ($7.75\cdot10^{-3}$) for a similar type of transition.

\begin{center}
  \begin{figure}[bt]
     \centering
      \includegraphics[width=\columnwidth]{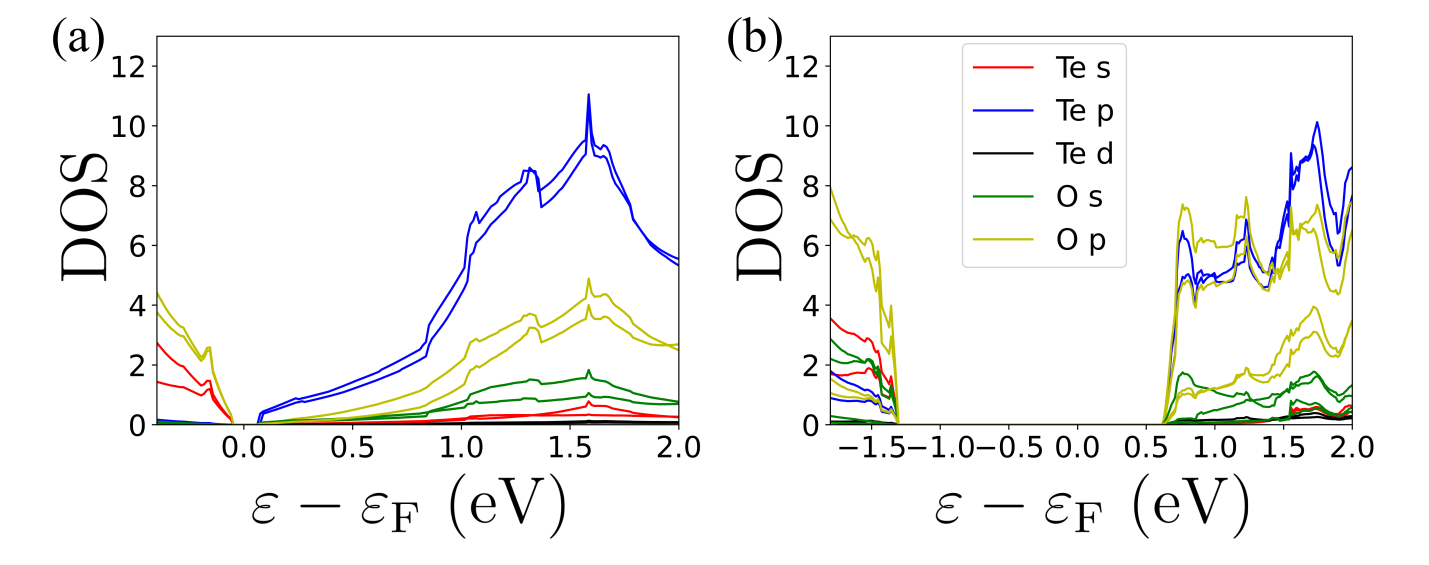}
      \caption{Calculated projected electronic density of states (DOS) of the TeO$_2$ crystal (a) in the high symmetry achiral rutile phase and (b) in the low symmetry chiral phase.} 
      \label{fig:PDOSTeO2} 
  \end{figure}
\end{center}

Interestingly, an analysis of the band structure and electronic density of states (see Fig.~\ref{fig:PDOSTeO2}) reveals that a significant change in the electronic structure also accompanies the structural phase transition. 
Indeed, the electronic band gap is substantially increased from a value of only $0.11$ eV in the high symmetry phase to a value of $1.93$ eV in the chiral structure contrary to the case of K$_3$NiO$_2$ where the band gap is essentially unaffected by the chiral transition~\cite{fava2025-KNO}. 
Therefore, inducing a chiral transition in this material would also tune its band gap by one order of magnitude, which might be interesting for applications. Clearly, this behavior is material dependent and cannot be extrapolated to other compounds emphasizing the need to evaluate these properties on a case-by-case basis.

This example illustrates how our methodology provides a clear pathway for identifying potential candidates for chiral displacive transitions in the case of TeO$_2$ crystal. 
We will now apply this approach to other types of compounds.

\subsection{The case of S\MakeLowercase{r}\textsubscript{2}A\MakeLowercase{s}\textsubscript{2}O\textsubscript{7}}
Here, we shall present another example of chiral phase transition predicted with the pseudosymmetry search algorithm. 
The Sr$_2$As$_2$O$_7$ compound exhibits experimentally a low symmetry phase of either $P4_1$ or $P4_3$ chiral space groups~\cite{Edhokkar_2012,Mbarek-13}. 
We could identify an achiral parent structure of $P4_2$ symmetry by applying the pseudosymmetry search algorithm. 
We have relaxed this new parent structure with a $k$-mesh of 4$\times$4$\times$2 and the obtained structural parameters can be found in the appendix.

%

\begin{figure*}[tb]
     \centering
      \includegraphics[width=12cm]{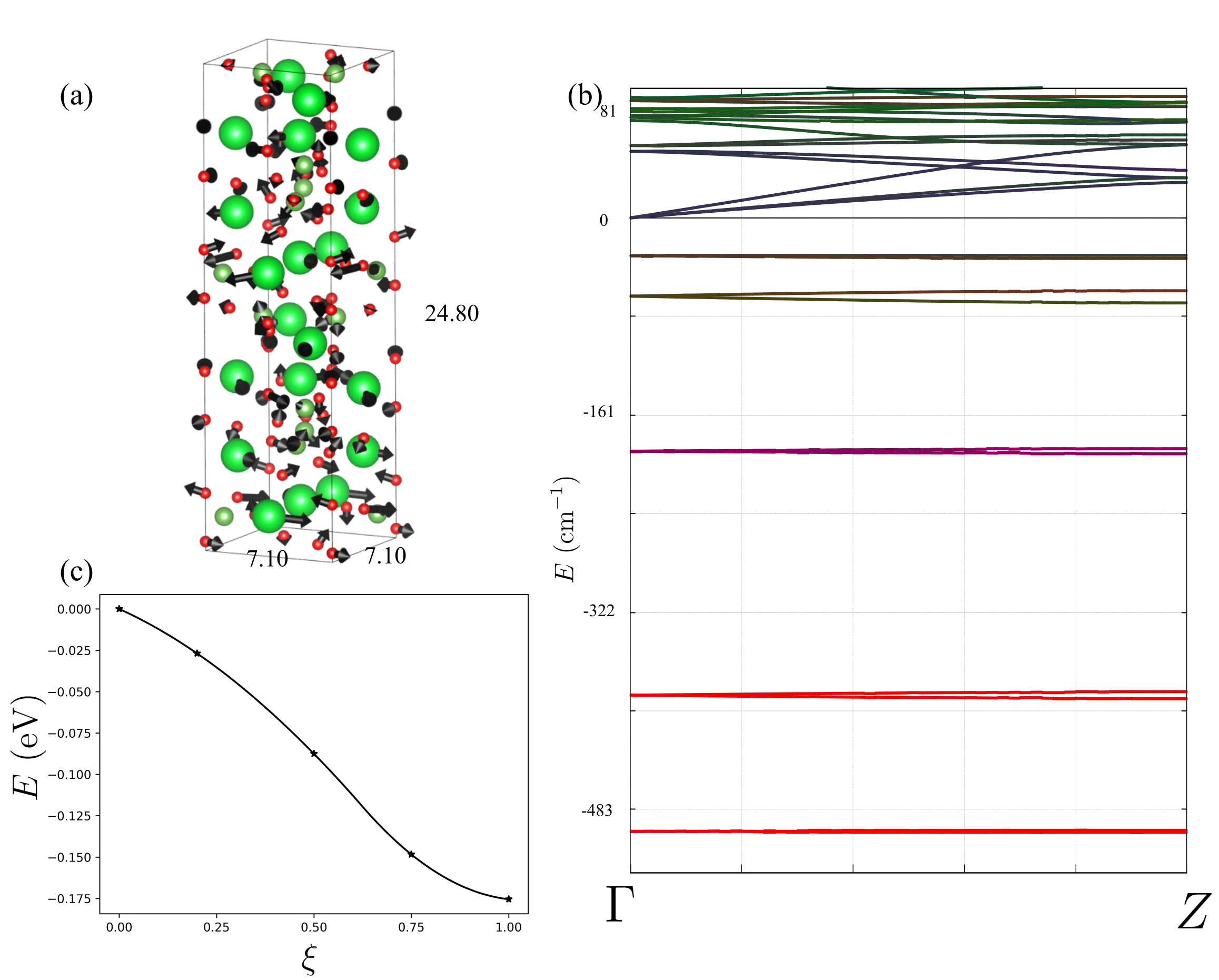}
      \caption{(a) Schematic representation of the Sr$_2$As$_2$O$_7$ crystal structure where atoms occupy the $P4_2$ high symmetry positions. 
      Arrows indicate the direction of the atom displacements that bring the system to the $P4_3$ chiral phase. 
      Green, yellow, and red balls represent Sr, As, and O atoms, respectively. (b) Calculated phonon band structure along the $\Gamma-Z$ path of the high symmetry phase of Sr$_2$As$_2$O$_7$. The color of the different branches are weighted according to the contribution of each chemical species to the dynamical matrix eigenvector (red for the O atom, green for the Sr and blue for the As atoms). (c) Energy surface when we condense the $Z_4$ unstable mode in normalized units.} 
      \label{fig:structureSr2As2O7} 
\end{figure*}

As in the previous case, a symmetry analysis of the supergroup–subgroup relationship reveals that this parent structure can be connected to the enantiomorphic space groups $P4_1$ and $P4_3$ through two different irreps at the $Z$ point. 
Indeed a SAM analysis gives $Z_3$ (with an amplitude of 0.054~\AA\ per ion) and $\Gamma_1$ (with an amplitude of 0.016~\AA\ per ion) for the $P4_1$ space group and $Z_4$ and $\Gamma_1$ for the $P4_3$ space group but with the same SAM amplitudes as $Z_3$ and $\Gamma_1$ in the $P4_1$ case.
As for TeO$_2$, the $\Gamma_1$ mode appears to be secondary while the $Z_3$ and $Z_4$ are the primary main modes.
Unlike K$_3$NiO$_2$ and TeO$_2$, where the transition to chiral enantiomorphic phases is driven by the same irreducible representation (irrep), the two chiral space groups in this case are associated with two distinct irreps. 
Our phonon calculations in the $P4_2$ achiral reference of Sr$_2$As$_2$O$_7$ are consistent with these results as two unstable modes are found, one with an imaginary frequency of $-503i$ cm$^{-1}$ and the $Z_3$ label and the other one with an imaginary frequency of $-501i$ cm$^{-1}$ and the $Z_4$ label as shown in Fig.~\ref{fig:structureSr2As2O7}.
These two modes of different symmetry have the same frequency, and condensing their respective eigenvectors and relaxing the structure gives the same gain of energy ($\Delta E=-0.18$ eV per atom), as we would expect for two pairs of enantiomorphic crystals. The substantial unstable frequencies and significant energy gain suggest that achieving experimental stability for the achiral parent structure may be challenging. However, these findings could motivate further experimental efforts to enhance the stability of this phase, potentially through the application of tensile strain or by optimizing growth conditions, as was successfully demonstrated for TeO$_2$~\cite{peng2022}.

The $Z_3$ and $Z_4$ modes involve mainly the layer-by-layer nonpolar in-plane and the polar out-of-plane movement of the atoms. In contrast, the $\Gamma_1$ involves mainly the atoms' polar in-plane and nonpolar out-of-plane movement. 
The composition of these movements generates a helical structure as shown in Fig.~\ref{fig:structureSr2As2O7}. 
By probing the helicity amplitude of the structure~\cite{gomez2024pros}, we found a value of $0.81\cdot10^{-3}$, with, as we would expect, a simple sign change between the $P4_1$ and $P4_3$ phases. 

A study of the electronic band structure reveals a substantial band gap increase accompanying the transition, similar to the case of TeO$_2$. 
The band gap is $0.95$ eV in the high-symmetry phase and $3.03$ eV in the low-symmetry phase.

This new example is interesting because it shows that our methodology can find displacive chiral phase transitions in which two different phonon modes are responsible for the two enantiomorphic space groups.
In the limit of our knowledge, such a possibility has not been previously discussed in the literature.
\subsection{The case of N\MakeLowercase{a}\textsubscript{2}S\MakeLowercase{e}O\textsubscript{9}}
We shall now discuss the case of Na$_2$SeO$_9$. 
This material is only reported in the Materials Project and with the $P4_122$ and $P4_322$ pair of enantiomorphic space groups~\cite{materialsproject_na2seo9}. 
While no experimental report exists for this crystal, DFT calculations predict it to be stable. Given the scarcity of reported crystals within the enantiomorphic space groups, we can use it as a new test case for our search method for displacive chiral phase transition. 
After applying the pseudosymmetry search and relaxing the structures with a $k$-mesh sampling of 4$\times$4$\times$4, we obtain a high symmetry parent structure with the space group $P4_2/mmc$ and with structural data reported in the appendix, that can be connected to the enantiomorphic space groups by phonon modes at the $Z$ point.

\begin{figure*}[tb]
     \centering
      \includegraphics[width=12cm]{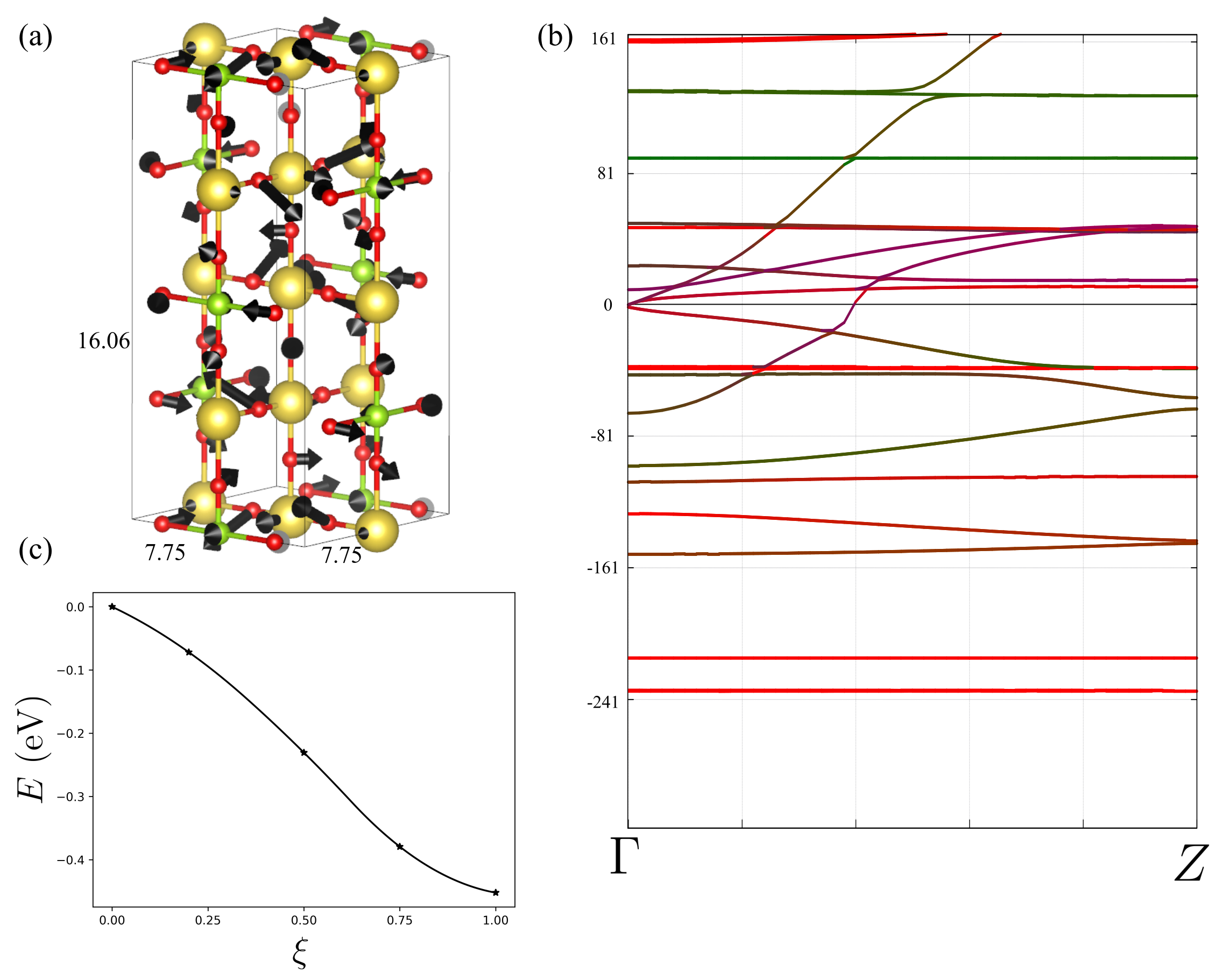}
      \caption{(a) Schematic representation of the Na$_2$SeO$_9$ crystal structure where atoms occupy the high symmetry positions ($P4_2/mmc$ space group). 
      Arrows indicate the direction of the atom displacements that bring the system to the $P4_322$ chiral phase. Gold, green, and red balls represent Na, Se, and O atoms. (b) Calculated phonon band structure along the $\Gamma-Z$ path of the high symmetry phase of Na$_2$SeO$_9$. The color of the different branches are weighted according to the contribution of each chemical species to the dynamical matrix eigenvector (red for the O atom, green for the Na and blue for the Se atoms). (c) Energy surface when we condense the $Z_3$ unstable mode in normalized units.} 
      \label{fig:structureNa2SeO9} 
\end{figure*}

By performing the phonon calculations at the $Z$-point, we found several unstable modes, as shown in Fig.~\ref{fig:structureNa2SeO9}.
The irreps of the two most unstable modes are $Z_1$ and $Z_2$, which gives a nonchiral phase transition to space group $Pmmm$ and $Pccm$ respectively, with a degenerate frequency of $-237$ cm$^{-1}$. 
The third and fourth modes correspond to the $Z_3$ and $Z_4$ irreps, which are also degenerate and drive a chiral phase transition showing the same frequency of $-217$ cm$^{-1}$ and similar symmetry. 
In contrast to TeO$_2$ or K$_3$NiO$_2$, diagonal distortion domains $(a,a)$ or $(-a,-a)$ in this system lead to a $Pmma$ achiral space group symmetry, whereas distortions along $(a,0)$ or $(0,a)$ drive the chiral transition towards the $P4_122$ and $P4_322$ space groups, respectively. 

Upon relaxing the previous unstable modes, we observe that, despite exhibiting the highest instability, the $Z_1$ and $Z_2$ modes do not lead to the most stable phase, as they provide an energy gain of $\Delta E_1=-0.007$ eV per atom versus $\Delta E_2=-0.45$ eV per atom for the $Z_3$ and $Z_4$ modes in the chiral phase. 
A SAM decomposition of the relaxed $Z_3$  mode in $(0,a)$ domain gives $\Gamma_1$ (with amplitude 0.15\AA per ion) and $Z_3$ (with amplitude 0.30\AA per ion). 
The relaxation of the $Z_3$ and $Z_4$ modes along the diagonal domains $(a,a)$ or $(-a,-a)$  with the achiral $Pmma$ phase gives an energy gain of $E_3=-0.18$ eV per atom, which is larger than the relaxation coming from the $Z_1$ and $Z_2$ modes but lower than the chiral phase.
Hence, Na$_2$SeO$_9$ has the potential to exhibit a displacive chiral phase transition.

Computing the handedness of the chiral structure, we obtain a value of $2.70\cdot10^{-2}$. 
This significant handedness, compared to that of TeO$_2$ or K$_3$NiO$_2$, is explained by the increased magnitude of the distortion, which is three times larger. 
The magnitude of the handedness amplitude is quadratic concerning the chiral distortion amplitude~\cite{gomez2024pros}, which explains the large helicity value obtained for Na$_2$SeO$_9$.
In Fig.~\ref{fig:structureNa2SeO9}, we give a schematic representation of the crystal helical distortion from the $P4_2/mmc$ phase towards the $P4_322$ phase.

Lastly, regarding the band structure, the electronic band gap is reduced from $0.61$ eV in the high symmetry phase to gap closure in the chiral phase. 
This contrasts the previous three examples, where the chiral distortion strongly increased the band gap.
\subsection{Other materials showing no displacive chiral transitions}

This last section will discuss additional crystal structures for which either no candidates for parent structures were found using a pseudosymmetry search approach or no unstable phonon modes driving the chiral transition could be identified.
\begin{figure}[tb]
     \centering
      \includegraphics[width=6cm]{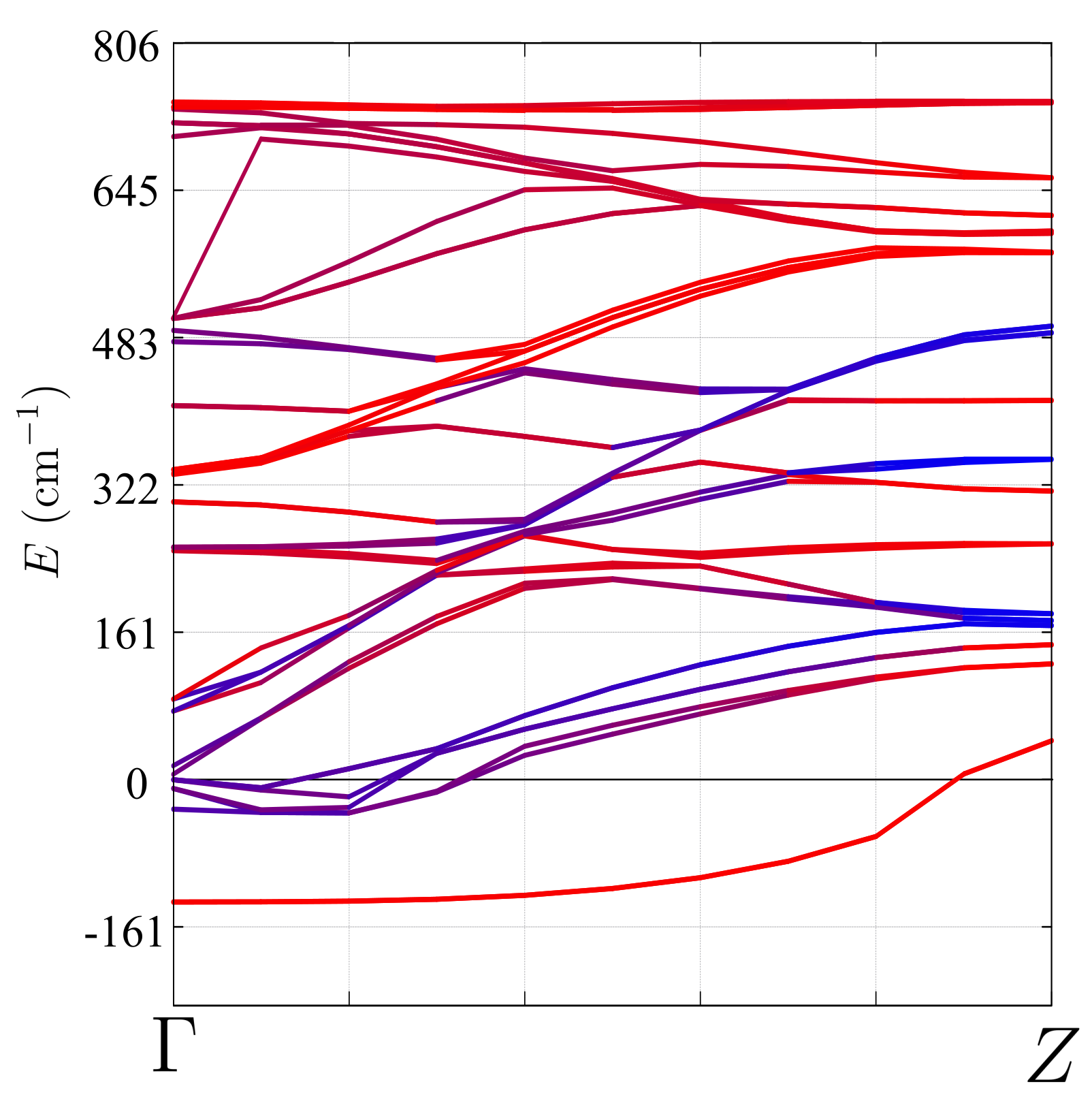}
      \caption{Calculated phonon band structure along the $\Gamma-Z$ path of the high symmetry phase of As$_2$O$_5$. The color of the different branches are weighted according to the contribution of each chemical species to the dynamical matrix eigenvector (red for the O atom, and blue for the As atoms).} 
      \label{fig:phonAs2O5} 
\end{figure}

First, let us consider the case of As$_2$O$_5$. 
The reported low symmetry chiral phase corresponds to a $P4_12_12$ or $P4_32_12$ space group~\cite{Jansen-79}, and we found a parent phase of symmetry $P4_2/mnm$ following a pseudosymmetry search approach. 
From symmetry considerations, we found that the irrep that can connect with the chiral phase should be observed at the $Z$ point. 
\begin{table*}[hbtp]
\caption{\label{tab:take_home}Summary table collecting chiral space groups, achiral parent structures found with pseudosymmetry, irreps, and corresponding domains driving the phase transition, frequency of the instabilities, energy gains and helicity. The~~\XSolidBrush~~sign indicates that either no parent phase was found or no soft phonon modes driving the transition were identified. Energy differences between high-symmetry and chiral phases after relaxation of the eigendisplacements are given per atom.}
\begin{tabular}{c  c  c  c  c  c  c}
\hline
   Compound& Chiral group &Achiral Ref.&Irrep&$\nu$ (cm$^{-1}$)&E (eV/atom) & $\mathcal{H}$\\ \hline
TeO$_2$ & \shortstack{$P4_12_12$\\$P4_32_12$}& $P4_2/mnm$& \shortstack{$Z_4~(a,a)$\\ $Z_4~(a,-a)$}&-370&-0.13&$7.38\cdot10^{-3}$\\
\hline
Sr$_2$As$_2$O$_7$ & \shortstack{$P4_1$\\$P4_3$}& $P4_2$& \shortstack{$Z_3$~$(a)$\\ $Z_4~(a)$}&\shortstack{-503\\-501}&-0.18&$0.81\cdot10^{-3}$\\
\hline
Na$_2$SeO$_9$ & \shortstack{$P4_122$\\$P4_322$}& $P4_2/mmc$& \shortstack{$Z_3$~$(a,0)$\\ $Z_3~(0,a)$}&-217&-0.45&$2.70\cdot10^{-2}$\\
\hline
As$_2$O$_5$ & \shortstack{$P4_12_12$\\$P4_32_12$}& $P4_2/mnm$& \shortstack{$Z_4~(a,a)$\\$Z_3~(a,a)$}&\XSolidBrush&\XSolidBrush&\XSolidBrush\\
\hline
Rb$_2$Be$_2$O$_3$ & \shortstack{$P4_12_12$\\$P4_32_12$}& \XSolidBrush& \XSolidBrush&\XSolidBrush&\XSolidBrush&\XSolidBrush\\
\hline
CaTe$_2$O$_3$ & \shortstack{$P4_12_12$\\$P4_32_12$}& \XSolidBrush& \XSolidBrush&\XSolidBrush&\XSolidBrush&\XSolidBrush\\
\hline
\end{tabular}
\end{table*}
However, unlike previous cases, no unstable mode is found when we compute the phonons at the $Z$ point as it is shown in Fig.~\ref{fig:phonAs2O5}.
This suggests that displacive chiral phase transitions in this system are not physically present, at least at the bulk level.

Another reason a material may not exhibit a displacive chiral phase transition is because it is impossible to identify any achiral parent structures through a pseudosymmetry search. 
We found that this is the case for the Rb$_2$Be$_2$O$_3$ compound that presents $P4_12_12$ or $P4_32_12$ low-symmetry chiral phases~\cite{materialsproject_rb2be2o3} or CaTeO$_3$ that shows $P4_1$ or $P4_3$ chiral space groups~\cite{Giester-09}. 
In this second situation, we have a clear symmetry demonstration that no achiral phase can be built from the chiral one.
\section{Discussion}
In this work, we have proposed a method combining pseudosymmetry search and first-principles calculations to identify achiral parent structures from known chiral phases connected through a soft phonon mode as present in displacive phase transitions. 
Our methodology, schematized in Fig.~\ref{fig:algorithm}, helps discover new candidates for such phase transitions and disregards those where such a transition is impossible. 
In Table~\ref{tab:take_home}, we summarize the different cases where we have applied our methodology collecting the most important results of the study.
The methodology helps to identify whether a soft phonon mode with the right symmetry can connect the achiral and chiral phases through continuous and relatively small displacements, e.g., TeO$_2$, Sr$_2$As$_2$O$_7$ or Na$_2$SeO$_9$.
It also helps to disregard the cases where the pseudosymmetry would predict a possible achiral-to-chiral connection according to the symmetry. Still, no unstable mode is observed in the achiral phase from the DFT calculation, so it is physically improbable that a displacive phase transition exists, e.g., in the case of As$_2$O$_5$.
Although, other mechanisms not captured by the present workflow, such as electronic instabilities like the Jahn-Teller one, could also potentially lead to chiral phase transitions~\cite{Isobe2002}.
Finally, it also helps to quickly show that no achiral phase can be constructed from the pseudosymmetry approach, which guarantees the absence of any smooth achiral to chiral transition as no high symmetry phase can be constructed, e.g., the cases of CaTeO$_3$ and Rb$_2$Be$_2$O$_3$. 
We could identify three new possible candidates (TeO$_2$, Sr$_2$As$_2$O$_7$ or Na$_2$SeO$_9$) that can have a soft phonon mode connecting the achiral and the chiral phase. This suggests that many more materials may display this behavior beyond the few examples reported in the literature so far.
Interestingly, Sr$_2$As$_2$O$_7$ appears to belong to a new class of systems where different irreps govern the chiral transitions to each of the enantiomorphic space groups. This discovery underscores the potential of our algorithm to reveal novel and intriguing chiral transitions. Further theoretical investigations would be valuable to determine whether similar behavior exists in other compounds, and to explore the possibility of additional new mechanisms for chiral transitions, thus opening exciting avenues for future research.

Our methodology provides a systematic and accurate way to screen various materials to search for new candidates for displacive chiral phase transitions.
As only one soft-mode driven chiral transition has been reported within the 22 enantiomorphic space groups, this methodology is well-suited to be implemented on a high-throughput scheme and check whether the lack of reported structures is due to an overlooked problem or if it is related to some chemical or physical reason~\cite{bousquet2025}.
We hope this work will inspire the community to explore new soft-mode-driven chiral phase transitions together with the non-linear phononics  ones~\cite{romao2024} and to motivate experimentalists to check our predictions.
\section{Methods}
\subsection{Computational details}
For all the calculations we converged the structural data of the parent structure with the {\sc{Abinit}} code~\cite{Gonze-20} (version 9.10.5) with a PBEsol exchange-correlation functional, following a plane wave-pseudopotential approach with optimized norm-conserving pseudopotentials as taken from the PseudoDojo server v4~\cite{hamann2013optimized,van2018pseudodojo} and an energy cutoff of $40$ Hartrees. 
Details about the $k$-mesh sampling employed specifically for each compound will be given for each case in respective sections. 
A value of $10^{-6}$ Har/Bohr was employed on the forces to stop the structural relaxations and a value of $10^{-7}$ Ha was used for the electronic residual self-consistent cycle stop. 
Phonon instabilities and Born effective charges were computed within the density functional perturbation theory framework as implemented in {\sc{abinit}}~\cite{Gonze1997,Gonze-20}.
\section{Data Availability}
All data regarding the structural properties of the high symmetry and low symmetry phases are provided in the appendix and in the dedicated supplementary file in plain text format.
\acknowledgments
We acknowledge helpful discussions with Luis Elcoro, administrator of the Bilbao Crystallographic Server.
F.G.O. acknowledges financial support from MSCA-PF 101148906 funded by the European Union and the Fonds de la Recherche Scientifique (FNRS) through the FNRS-CR 1.B.227.25F grant. 
F.G.-O. and E.B. acknowledge the Fonds de la Recherche Scientifique (FNRS) for financial support, the PDR project CHRYSALID No.40003544 and the Consortium des Équipements de Calcul Intensif (CÉCI), funded by the F.R.S.-FNRS under Grant No. 2.5020.11 and the Tier-1 Lucia supercomputer of the Walloon Region, infrastructure financed by the Walloon Region under the grant agreement No. 1910247. 
F.G.-O. and E. B. also acknowledge support from the European Union’s Horizon 2020 research and innovation program under Grant Agreement No. 964931 (TSAR). 
We also thank the Pittsburgh Supercomputer Center (Bridges2) and San Diego Supercomputer Center (Expanse) through allocation DMR140031 from the Advanced Cyberinfrastructure Coordination Ecosystem: Services \& Support (ACCESS) program, which National Science Foundation supports grants \#2138259, \#2138286, \#2138307, \#2137603, and \#2138296. 
We also recognize the computational resources provided by the WVU Research Computing Dolly Sods HPC cluster, which is funded in part by NSF OAC-2117575.
We also recognize the support of the West Virginia High Education Policy Commission under the call Research Challenge Grand Program 2022, Award RCG 23-007 and NASA EPSCoR Award 80NSSC22M0173
\appendix
\section{Structural data}
In this appendix we report the structural data of the high-symmetry phases encountered for the TeO$_2$, Na$_2$SeO$_9$ and Sr$_2$As$_2$O$_7$ crystals together with their associated displacements towards their respective chiral phases as reported in the main text.
\begin{table}[hb]
\caption{\label{tab:dataTeO2}Structural data in reduced coordinates for the high symmetry parent phase of TeO$_2$ found with the pseudosymmetry search algorithm and after DFT relaxation. 
Unit cell parameters for the tetragonal structure are $a=b=5.060$\AA, $c=3.401$\AA. 
Forces and stresses where converged up to $10^{-6}$ Har/Bohr ($5.14\cdot 10^{-5}$ eV/\AA). 
The last three columns represent the unrelaxed eigenvector displacements following the $Z_3$ mode towards space group number 92 ($P4_12_12$) in the home unit cell.}
\begin{tabular}{c  c  c  c c c c}
\hline
   Atom&x&y&z&$\delta_x$&$\delta_y$&$\delta_z$\\ \hline
Te &0.0000 & 0.0000 & 0.0000&-0.0204&-0.0204&0.0000\\
Te&0.5000  &0.5000  &0.5000&0.0204&-0.0204&0.0000\\
O & 0.6976 & 0.3024 & 0.0000&-0.0417&-0.0372&0.0419\\
O & 0.3024 & 0.6976 & 0.0000&-0.0372&-0.0417&-0.0419\\
O & 0.8024 & 0.8024& 0.5000&0.0417&-0.0372&-0.0419\\
O & 0.1976& 0.1976 & 0.50000&0.0372&-0.0417&0.0419\\
\hline
\end{tabular}
\end{table}
\begin{table}[hb]
\caption{\label{tab:dataNa2SeO9}Structural data in reduced coordinates for the high symmetry parent phase of Na$_2$SeO$_9$ found with the pseudosymmetry search algorithm after relaxation in the low symmetry basis. Unit cell parameters for the tetragonal structure are $a=b=7.755$\AA, $c=16.079$\AA. Forces and stresses where converged up to $10^{-6}$ Har/Bohr ($5.14\cdot 10^{-5}$ eV/\AA). The last three columns represent the displacement following the chiral mode towards space group 95 ($P4_322$).}
\begin{tabular}{c  c  c  c  c c c c}
\hline
   Atom&Wyckoff &x&y&z&$\delta_x$&$\delta_y$&$\delta_z$\\ \hline
Na &  4b & 0.5000&  0.5000&  0.0000&0.0000&0.1447&0.0000\\
Na &  4a & 0.0000&  0.5000&  0.5000&0.0000&-0.0322&0.0000\\
Se &  4b  &0.5000&  0.0000&  0.0000&0.0000&0.0474&0.0000\\
O  &  4c & 0.5000&  0.5000&  0.6250&0.0749&0.0749&0.0000\\
O  &  8d & 0.0000&  0.1753&  0.2500&-0.1075&0.0597&0.0109 \\
O  &  8d & 0.0000&  0.5000&  0.1471&-0.0814&-0.0164&-0.0206\\
O  &  8d & 0.2808&  0.1448&  0.5000&-0.0416&-0.2538&-0.0589\\
O  &  8d & 0.2808&  0.1448&  0.0000&0.1342 &-0.1193&-0.0131\\
\hline
\end{tabular}
\end{table}
\begin{table}[hb]
\caption{\label{tab:dataSr2As2O7}Structural data in reduced coordinates for the high symmetry parent phase of Sr$_2$As$_2$O$_7$ found with the pseudosymmetry search algorithm after relaxation expressed in the low symmetry basis. 
Unit cell parameters are $a=b=7.098$\AA, $c=24.805$\AA. 
Forces and stresses where converged up to $10^{-6}$ Har/Bohr ($5.14\cdot 10^{-5}$ eV/\AA). 
The last three columns represent the unrelaxed eigenvector displacements following the $Z_4$ mode towards space group number 78 ($P4_3$).}
\begin{tabular}{c  c  c  c  c c c c}
\hline
   Atom&Wyckoff &x&y&z&$\delta_x$&$\delta_y$&$\delta_z$\\ \hline
Sr& 4a&0.2366&0.8626&0.1870&-0.0154&0.0674&-0.0019\\
Sr& 4a&0.7634&0.1374&0.1870&0.0331&0.0558&-0.0008\\
Sr& 4a&0.5000&0.0000&0.3327&0.1017&0.0745&-0.0057\\
Sr& 4a&0.5000&0.5000&0.3388&-0.0698&0.0536&0.0008 \\
As& 4a&0.4096&0.7431&0.4679&0.0366&0.0222&-0.0139\\
As& 4a&0.5904&0.2570&0.4679&0.0314&-0.0023&0.0135 \\
As& 4a&0.2259&0.9672&0.0595&0.0081&0.0184&0.0029  \\
As& 4a&0.7741&0.0328&0.0595&-0.0250&-0.0025&0.0027 \\
O & 4a&0.3209&0.5886&0.2583&-0.0001&0.0068&-0.0073  \\
O & 4a&0.6791&0.4114&0.2583&0.0281&0.0238&0.0144    \\   
O & 4a&0.5084&0.7665&0.4052&0.0566&0.0528&-0.0101  \\  
O & 4a&0.4916&0.2335&0.4052&0.0424&0.0477&0.0076   \\
O & 4a&0.3123&0.9583&0.4840&0.0032&-0.0029&-0.0029 \\ 
O & 4a&0.6877&0.0418&0.4840&-0.0029&-0.0357&0.0019\\
O & 4a&0.5000&0.0000&0.2297&0.0529&-0.0040&-0.0130\\
O & 4a&0.0000&0.5000&0.3343&0.0823&0.1301&-0.0057   \\      
O & 4a&0.0000&0.0000&0.0187&0.0106&0.0772&-0.0015\\
O & 4a&0.2355&0.7281&0.0678&-0.0006&0.0011&-0.0172 \\ 
O & 4a&0.7645&0.2720&0.0678&0.0083&-0.0117&0.0154\\
O & 4a&0.2224&0.5993&0.4499&0.0806&-0.0185&-0.0046 \\ 
O & 4a&0.7776&0.4007&0.4499&0.0528&-0.0082&0.0181  \\
O & 4a&0.0000&0.0000&0.1182&0.0343&-0.0665&0.0073\\
\hline
\end{tabular}
\end{table}
\clearpage
\newpage
\newpage
\end{document}